\documentclass[conference]{IEEEtran}
\IEEEoverridecommandlockouts
\usepackage{cite}
\usepackage{amsmath,amssymb,amsfonts}
\usepackage{algorithmic}
\usepackage{graphicx}
\usepackage{textcomp}
\usepackage{xcolor}
\def\BibTeX{{\rm B\kern-.05em{\sc i\kern-.025em b}\kern-.08em
    T\kern-.1667em\lower.7ex\hbox{E}\kern-.125emX}}
\usepackage{hyperref}
\begin{document}

\title{GANs for EVT Based Model Parameter Estimation in Real-time Ultra-Reliable Communication

\author{Parmida Valiahdi and Sinem Coleri, \IEEEmembership{Fellow,~IEEE}\vspace*{-\baselineskip}
\thanks{Parmida Valiahdi and S. Coleri are with the Department of Electrical and
Electronics Engineering, Koc University, Istanbul, e-mail: \{pvaliahdi23, scoleri\}@
ku.edu.tr. Sinem Coleri acknowledges the support of the Scientific and Technological
Research Council of Turkey 2247-A National Leaders Research
Grant \#121C314.}}

}

\maketitle

\begin{abstract}
The Ultra-Reliable Low-Latency Communications (URLLC) paradigm in sixth-generation (6G) systems heavily relies on precise channel modeling, especially when dealing with rare and extreme events within wireless communication channels. This paper explores a novel methodology integrating Extreme Value Theory (EVT) and Generative Adversarial Networks (GANs) to achieve the precise channel modeling in real-time. The proposed  approach harnesses EVT by employing the Generalized Pareto Distribution (GPD) to model the distribution of extreme events. Subsequently, Generative Adversarial Networks (GANs) are employed to estimate the parameters of the GPD. In contrast to conventional GAN configurations that focus on estimating the overall distribution, the proposed approach involves the incorporation of an additional block within the GAN structure. This specific augmentation is designed with the explicit purpose of directly estimating the parameters of the Generalized Pareto Distribution (GPD). Through extensive simulations across different sample sizes, the proposed GAN based approach consistently demonstrates superior adaptability, surpassing Maximum Likelihood Estimation (MLE), particularly in scenarios with limited sample sizes.
\end{abstract}

\begin{IEEEkeywords}
6G, ultra-reliable low-latency communications (URLLC), extreme value theory (EVT), Generalized Pareto Distribution (GPD), parameter estimation, Generative Adversarial Networks (GANs), machine learning, wireless channel modeling.
\end{IEEEkeywords}

\section{Introduction}
Ultra-Reliable Low-Latency Communications (URLLC) is one of the most transformative features of the sixth generation (6G) networks, effectively meeting the escalating demands of real-time control, interactions, and decision-making across diverse applications, such as remote surgeries, industrial automation and autonomous vehicular platoons. URLLC ensures the delivery of critical information with a packet error rate in the range of $10^{-9}$ - $10^{-5}$, and delays on the order of milliseconds \cite{b4}. Achieving ultra-high reliability and low latency at this level necessitates communication systems to incorporate an accurate channel modeling based on the derivation of the tail statistics.

In the realm of URLLC channel modeling, three distinct approaches have emerged. The first approach is based on the extrapolation of the existing channel models to the ultra-reliable region \cite{b11}. The second approach challenges and redefines conventional concepts, such as coherence time, to meet the unique demands of ultra-reliable communication \cite{b12,b13}. Additionally, alternative metrics for evaluating channel reliability, like averaged reliability for dynamic scenarios and probably correct reliability for static conditions, have been proposed \cite{b14}. The third approach is based on the usage of Extreme Value Theory (EVT) to analyze rare data instances corresponding to the tail statistics of probability distributions \cite{b1,b2,b3,b7,b21,b22}. EVT employs the Generalized Pareto Distribution (GPD) to estimate the tail statistics of the wireless channel, which often involve extremely high interference or deep fading. These papers employ Maximum Likelihood Estimation (MLE) to determine the GPD parameters, assuming the availability of a large amount of data. However, integrating EVT based statistics into real-time communication requires the usage of more efficient estimation techniques to achieve high accuracy and adaptability with limited data acquired within low latency.

In the domain of channel modeling, machine learning techniques have gained prominence for accurately estimating parameters even when working with limited data \cite{b9,b10,b16,b17,b18,b19,b20}. \cite{b16} proposes a Convolutional Neural Network (CNN) architecture to estimate the parameters of fading channel, including fading rate and amplitude, outperforming traditional statistical methods and manual feature engineering. In \cite{b18}, a neural network structure is employed to estimate the parameters of Rayleigh fading channel, improving the estimation robustness and accuracy when compared to traditional  Maximum Likelihood Detector (MLD). In \cite{b9}, Generative Adversarial Networks (GANs) are utilized for real-time estimation of the parameters, such as signal-to-noise ratio (SNR), channel gain, and channel impulse response, in AWGN channels.  \cite{b10} integrates GANs and LSTM networks to improve the accuracy of predicting future channel behavior, presenting a fundamental advancement over conventional approaches. \cite{b19} proposes synthetic data generation using GANs to estimate the distribution of rare events in URLLC wireless networks. Despite the extensive utilization of machine learning methods in communication networks, these techniques have not been directly employed to estimate parameters of the GPD for ultra-reliable communication.

In this paper, we introduce a novel methodology for the estimation of the GPD parameters based on the usage of GANs for real-time URLLC communication. Unlike previous applications of GANs in channel estimation, which primarily focused on directly estimating the overall channel distribution, our approach leverages GANs to estimate the GPD parameters. The performance of the proposed methodology is evaluated across various data sample size compared to the MLE based approach. 

The rest of the paper is organized as follows. Section II describes the system model and the background on EVT-based GPD modeling and generating probability distributions with GANs. Section III introduces the proposed GAN-based algorithm for estimating GPD parameters. Section IV provides the performance evaluation of the proposed methodology compared to MLE. Finally, Section V summarizes our findings and outlines potential directions for future research.

\section{System Model}
We consider a real-time ultra-reliable communication system that encounters severe fading, leading to remarkably low received power levels. In this scenario, the transmitter (Tx) sends a packet to a receiver (Rx) at constant transmit power over an unknown channel. Therefore, utilizing received signal power is akin to utilizing the squared amplitude of the channel state information. The rare events corresponding to severe fading conditions are modeled by GPD based on EVT. The receiver employs GANs to estimate the parameters of the GPD fitted to the channel tail distribution. 

We assume that the distribution of received powers is stationary, meaning the parameters defining the GPD distribution remain constant. If non-stationarity is detected through the Augmented Dickey-Fuller (ADF) test, external factors influencing the GPD parameters are identified. The GPD parameters are then estimated for the sequence over which the channel can be considered stable, as explained in \cite{b23}. Our system comprises two fundamental components: EVT-based GPD modeling and Generating Probability Distributions with GANs.

\subsection{EVT-Based GPD Modeling}
We begin by utilizing EVT to model rare and extreme events in wireless communication channels. EVT provides a robust framework for analyzing the statistics of these extreme events, by modeling the probabilistic distribution of the values exceeding a predefined threshold. We assign a GPD to represent the tail behavior of these extreme events based on Theorem 4.1 in \cite{b7}. 

Consider a sequence of received powers, denoted by $X^n= {X_1, . . ., X_n}$, comprising independent and identically distributed (i.i.d.) stationary observations. For low enough threshold $u$, the probabilistic distribution of power values exceeding $u$, i.e., Pr$\{u-X|X<u\}$, is approximated by the GPD as given by\\
\begin{equation}
F_u(y;\sigma,\zeta) = 
1 - (1 + \frac{\zeta y}{\tilde{\sigma_u}})^{-\frac{1}{\zeta}},
\end{equation}
where $y$ represents a non-negative value denoting the exceedance below threshold $u$; $\zeta$ and $\tilde{\sigma_u} = (\sigma + \zeta(u - \mu))$ are the shape and modified scale parameters of the GPD, respectively; $\mu$ and $\sigma$ are the location and scale parameters of the generalized extreme value (GEV) distribution fitted to the Cumulative Distribution Function (CDF) of $M_n = max\{X_1, . . ., X_n\}$, respectively.

\subsection{Generating Probability Distributions with GANs}
GAN is a machine learning method consisting of two neural networks: one is the generator and the other is the discriminator. These two networks are trained together using adversarial training techniques. The generator creates synthetic data samples that resemble a specific distribution, often starting from random noise. It does so by learning from a real dataset and attempting to generate data that is indistinguishable from it. In contrast, the discriminator's task is to distinguish between real data and the synthetic data generated by the generator. The two components engage in a competitive process where the generator continuously refines its data generation process to produce samples that are increasingly realistic. Simultaneously, the discriminator improves its ability to differentiate between real and synthetic data. As a result, this adversarial interplay leads to the generator creating data that becomes progressively difficult to distinguish from real data. This enables GAN to determine the best fitting distribution to the real data.

\section{GAN Based GPD Parameter Estimation}

\begin{figure}[htbp]
\centerline{\includegraphics[width=90mm]{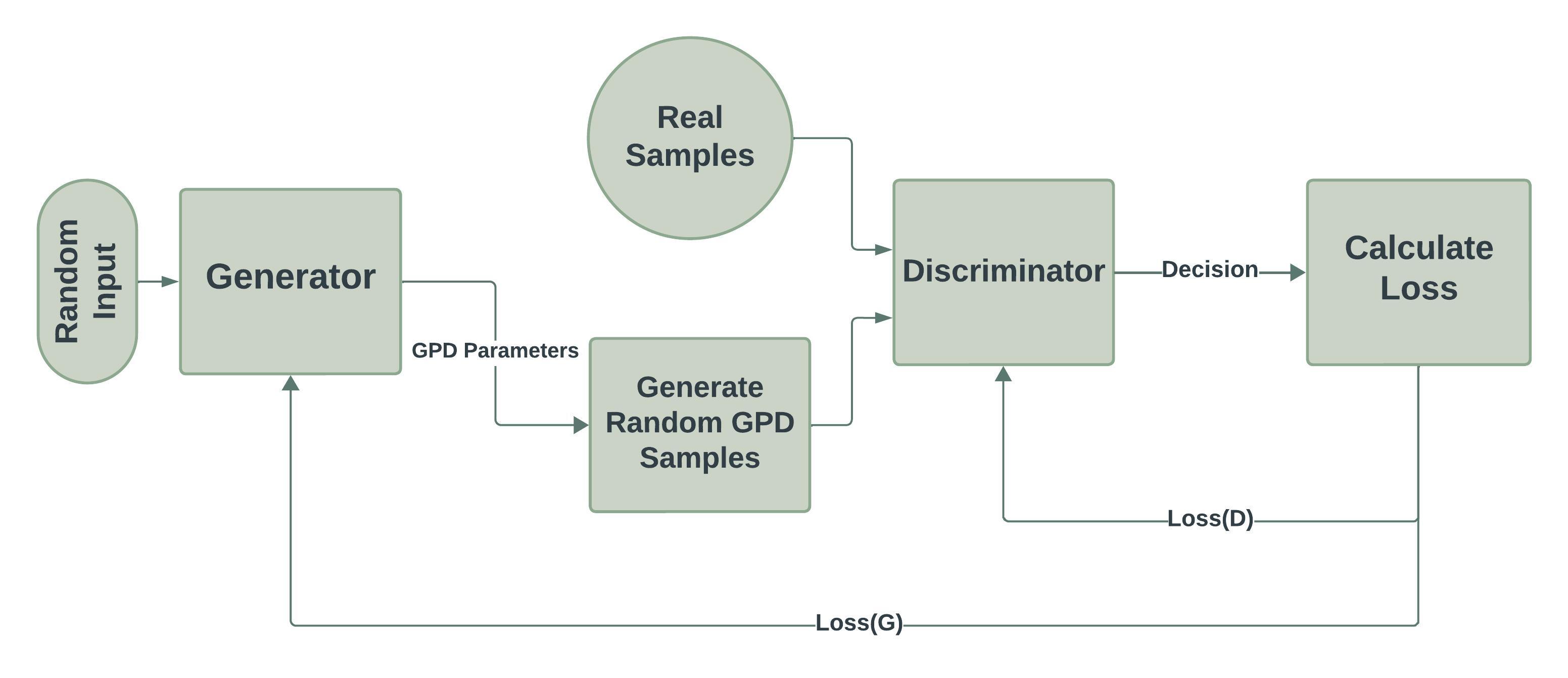}}
\caption{Generative Adversarial Network (GAN) Architecture for GPD Parameter Estimation}
\label{GAN}
\end{figure}

We propose a novel GAN based methodology to estimate the GPD parameters for extreme events in URLLC wireless channels. Our GAN architecture is a two-network system comprising a generator and a discriminator engaged in a competitive learning process, as shown in Fig. \ref{GAN}. This architecture is adapted for real-time updates, where the discriminator evaluates one sample at a time. The input to the generator consists of GPD random samples. Unlike traditional GAN setups that estimate the overall distribution, we have introduced an additional block to the GAN structure with the specific purpose of directly estimating the parameters of the GPD.

In this augmented architecture, the generator takes a set of random samples from the GPD distribution as input and  produce a two-dimensional vector representing the shape and scale parameters. Instead of using any random samples, we choose to generate GPD samples based on the mean and variance of the real dataset to speed up the learning process \cite{b24}. Although this estimation might be less reliable for datasets with fewer than 50 samples, it is a practical way to input the generator and streamline the training process. Additionally, we use these estimated parameters to set the initial weights of the generator, aligning them with the presumed shape and scale parameters that fit the GPD distribution of real data. The discriminator, responsible for distinguishing between real and generated data, is a binary classifier with a single output. Both the generator and discriminator architectures consist of multiple linear layers. During training, we employ a Binary Cross Entropy (BCE) loss function to guide the generator toward producing outputs that closely match the target GPD parameters. The GAN is trained iteratively with a combination of stochastic gradient descent (SGD) for both the generator and discriminator. Additionally, the convergence and performance of the GAN are continuously monitored through visualizations, including histograms of the generated and real data, and error plots across training epochs. 

\section{Performance Evaluation}
The goal of our simulations is to assess the efficacy of the proposed GAN-based GPD parameter estimation in comparison to the traditional MLE method across diverse sample sizes. 

\begin{figure}[htbp]
\centerline{\includegraphics[width=55mm, height=50mm]{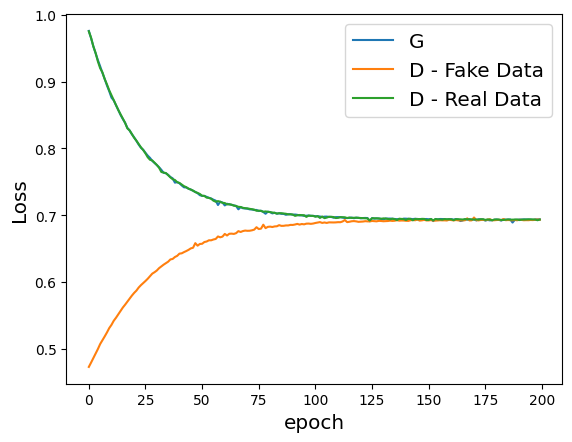}}
\caption{Loss Evolution for Generator and Discriminator over Epochs}
\label{Loss}
\end{figure}

The generator and discriminator of the proposed GAN based architecture consist of three layers, each equipped with 10 neurons. Employing the sigmoid activation function for the discriminator ensures the mapping of output between 0 and 1, which is apt for binary classification. For the generator, we utilize tanh activation exclusively for the shape parameter output, considering the GPD shape parameters should be within $[-1,1]$ range. Over a 200-epoch training period, weights of both networks are updated in 50 time steps per epoch. At each step, the generator processes 1000 GPD-distributed random samples, estimating shape and scale parameters, while the discriminator evaluates one real data sample and one generated based on the estimated parameters. Our training methodology involves utilizing various datasets, each generated with known shape and scale, of varying sample sizes. This approach ensures the generator learns parameter estimation based on dataset characteristics, irrespective of size. The three loss functions—generator loss (evaluated through Binary Cross Entropy), discriminator fake loss (quantified by BCE for fake data), and discriminator real loss (determined using BCE for real data)—are integral to our simulation framework. Optimal convergence is achieved when the generator loss aligns with discriminator decisions favoring fake data, the discriminator fake loss approaches discrimination against fake data, and the discriminator real loss converges towards perfect identification of real data. This stringent criterion ensures the effective training of our GAN model. Fig. \ref{Loss} visually represents our GAN's well-structured convergence towards the optimal loss.

GPD-distributed datasets are generated with varying sample sizes, each having known shape and scale parameters. Subsequently, for each sample size, we obtain shape and scale parameter estimates from both MLE and GAN. These estimates are first compared against the known true values by using the quantile-quantile (q-q) plot. The q-q plot is a graphical method used to compare the distribution of generated data to an idealized theoretical distribution, typically a straight line. When the q-q plot’s assigned line is the 45-degree line (representing a perfect match), it indicates that the estimation of parameters aligns well with the expected distribution. To further evaluate and compare these estimations, we utilize the slope error, calculated as $(1 - M)$, in which $M$ is the line slope, for each line. A smaller slope error, closer to zero, signifies a superior estimation in capturing the distribution characteristics. Given our knowledge of the actual parameters, we proceed to compare the estimated Probability Density Functions (PDFs) with the ground truth PDF. This comparison across different sample sizes provides valuable insights into the accuracy of parameter estimation.

\begin{figure}[h]
\centerline{\includegraphics[width=90mm]{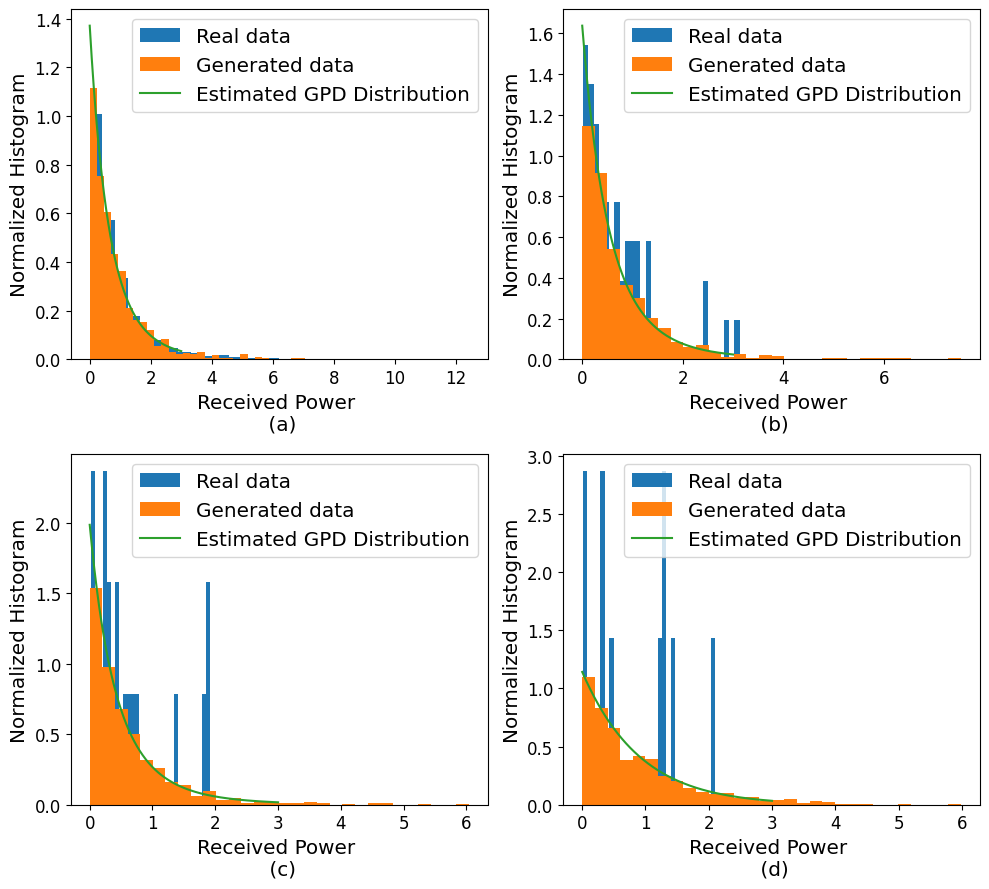}}
\caption{GPD parameter estimation of the proposed GAN based method for (a) 2000, (b) 50, (c) 20, and (d) 10 samples.}
\label{Est}
\end{figure}

Fig. \ref{Est} provides a comprehensive visualization of the GPD parameter estimations generated by the proposed GAN based approach across various sample sizes. These plots offer valuable insights into how our GAN-based approach performs as sample sizes decrease. Note that the task of accurately estimating GPD parameters is very challenging for small sample sizes since this limited number of samples may not fully represent the underlying GPD characteristics. The proposed GAN based approach demonstrates remarkable resilience and effectiveness even when the sample size is reduced to as few as 10 samples. 

\begin{figure}[h]
\centerline{\includegraphics[width=90mm]{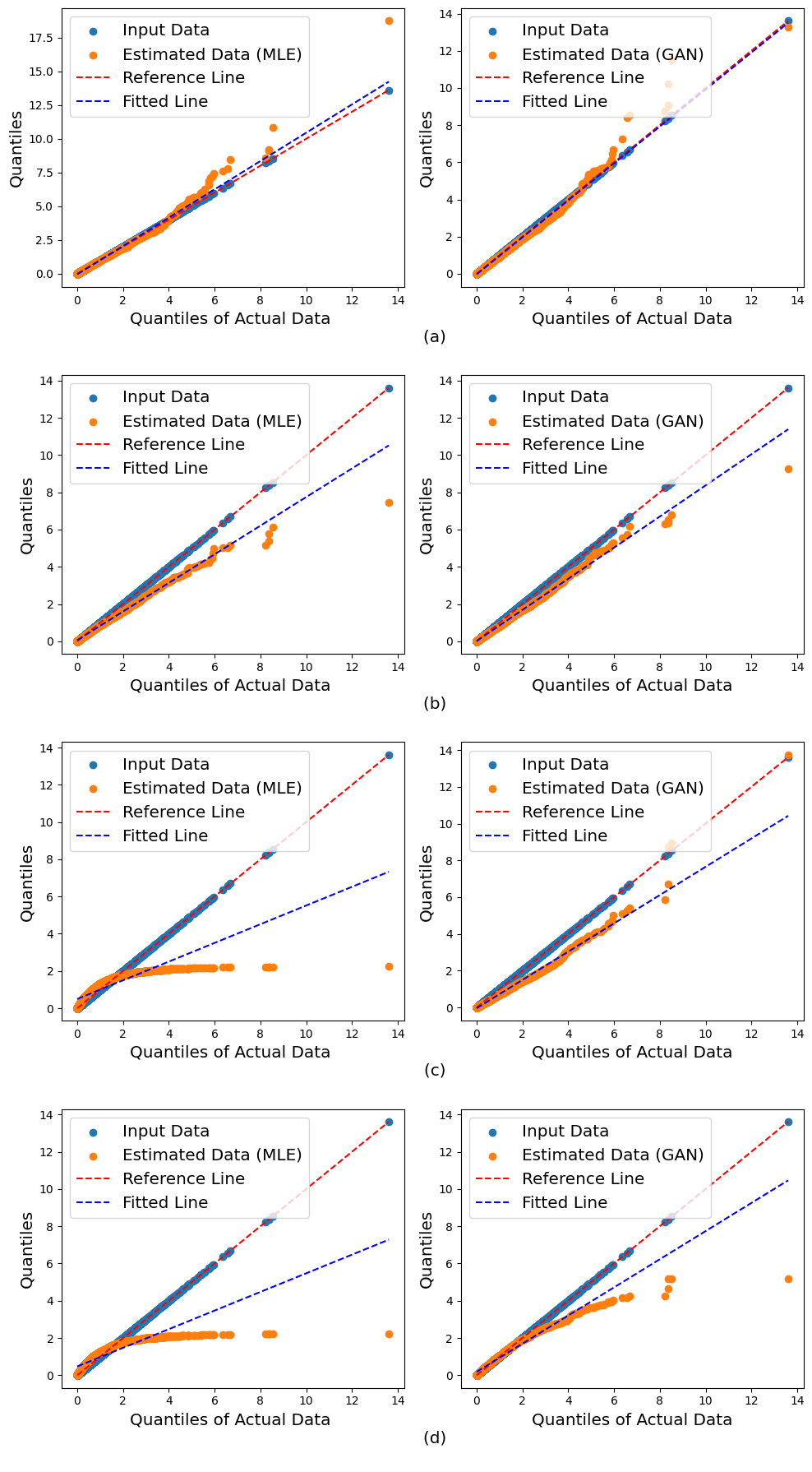}}
\caption{q-q Plots Comparing GAN and MLE Estimations for (a) 2000, (b) 50, (c) 20, and (d) 10 samples.}
\label{qq}
\end{figure}

Fig. \ref{qq} shows the q-q plots of the proposed GAN-based and MLE-based approaches for different sample sizes. The slope error is 0.0398, 0.2496, 0.4932 and 0.4916 by MLE for 2000, 50, 20 and 10 samples, respectively. On the other hand, the slope error is 0.0036, 0.1352, 0.2097 and 0.1930 by GAN for 2000, 50, 20 and 10 samples, respectively. For large sample sizes, both methods provide accurate estimations. GAN slightly outperforms MLE in terms of the accuracy of the estimated GPD parameters. However, as the sample size decreases to 50, both methods show some decrease in accuracy, while the GAN-based estimation performs notably better. As the sample size dwindles to less than 20 samples, MLE struggles to provide reliable estimates, while GAN continues to deliver an acceptable level of accuracy. This behavior underscores the robustness and adaptability of the GAN-based approach, particularly in situations with limited data, a crucial factor in real-time URLLC applications.

\begin{figure}[h]
\centerline{\includegraphics[width=90mm]{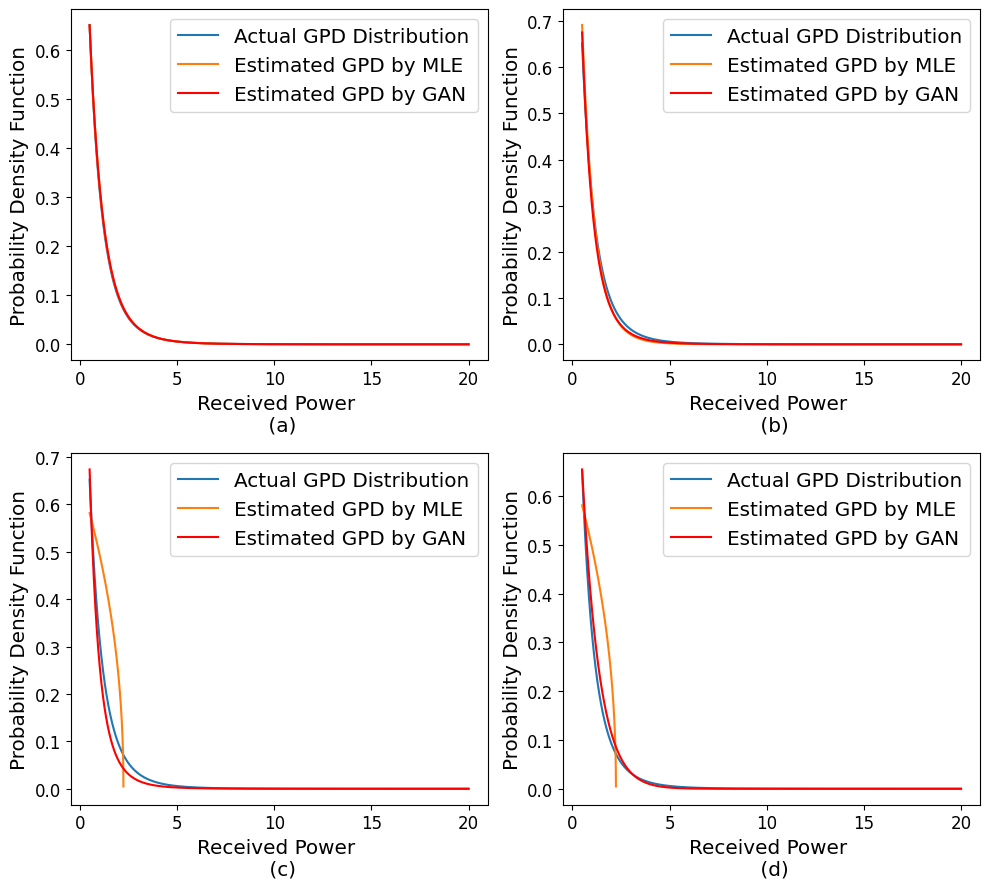}}
\caption{PDF comparison of MLE and proposed GAN based approach for (a) 2000, (b) 50, (c) 20, and (d) 10 samples.}
\label{pdf}
\end{figure}

Fig. \ref{pdf} offers a visual comparison of PDFs derived from estimations produced by the proposed GAN-based and MLE methods to the actual GPDs across varying sample sizes. GAN consistently outperforms MLE in terms of PDF accuracy, particularly when dealing with limited sample sizes. For large sample sizes, such as 2000 and 50, both the GAN and MLE estimations are accurate, with GAN exhibiting a slightly superior performance. However, as the sample size decreases to 20 and 10, the MLE estimation struggles to converge, resulting in less accurate estimations. In contrast, the GAN-based approach maintains its accuracy even with limited samples, demonstrating its robustness and effectiveness in scenarios with constrained data availability.

\section{Conclusion}
In this paper, we present a novel system model that leverages EVT for GPD modeling and utilizes GANs for real-time parameter estimation. This approach combines the robustness of EVT-based GPD modeling with the adaptability of GANs, thus offering an effective solution for the challenges posed by extreme events in wireless channels. Extensive simulations featuring q-q plots and PDF comparisons demonstrate the unique adaptability of GANs in effective parameter estimation across datasets of varying sizes. Notably, GANs excelled in cases where MLE struggled due to limited sample sizes, making them a particularly suitable choice for real-time URLLC applications. In the future, we plan to further extend the proposed GAN based approach by incorporating mechanisms for optimum threshold selection and validating the performance over actual experimental data.

\nocite{*}
\bibliographystyle{IEEEtran}
\bibliography{main}

\vspace{12pt}

\end{document}